# Direct Calculation of the Spin Stiffness in the $J_1$–$J_2$ Heisenberg Antiferromagnet

T. Einarsson and H. J. Schulz

*Laboratoire de Physique des Solides*[*]
*Université Paris-Sud*
*F-91405 Orsay, France*

(Oct. 25, 1994)

We calculate the spin stiffness $\rho_s$ for the frustrated spin-$\frac{1}{2}$ Heisenberg antiferromagnet on a square lattice by exact diagonalizations on finite clusters of up to 36 sites followed by extrapolations to the thermodynamic limit. For the non-frustrated case, we find that $\rho_s = (0.183 \pm 0.003) J_1$, in excellent agreement with the best results obtained by other means. Turning on frustration, the extrapolated stiffness vanishes for $0.4 \lesssim J_2/J_1 \lesssim 0.6$. In this intermediate region, the finite-size scaling works poorly – an additional sign that their is neither Néel nor collinear magnetic order. Using a hydrodynamic relation, and previous results for the transverse susceptibility, we also estimate the spin-wave velocity in the Néel-ordered region.

The question of the existence of long–range magnetic order (LRMO) in systems with frustrated interactions and strong (quantum or thermal) fluctuations is often difficult to decide. The traditional way of answering this question is by calculating magnetic order parameters. An alternative way is to consider the spin stiffness $\rho_s$, which is non-zero in a LRMO state. The stiffness has the advantage of being unbiased with respect to the order parameter, and constitutes, together with the spin-wave velocity, the fundamental parameter that determines the low-energy dynamics of magnetic systems [1]. It is therefore of importance to find accurate values for $\rho_s$.

The spin stiffness measures the energy cost to introduce a twist $\theta$ of the direction of spin between every pair of neighboring rows,

$$\rho_s = \left. \frac{d^2}{d\theta^2} \frac{E_0(\theta)}{N} \right|_{\theta=0} , \qquad (1)$$

where $E_0(\theta)$ is the ground-state energy as a function of the imposed twist, and $N$ is the number of sites. In the thermodynamic limit, a positive value of $\rho_s$ means that LRMO persists in the system, while a zero value reveals that there is no LRMO, as is the case in a spin-liquid. When looking at a finite system, things are more complicated. Here the stiffness is only zero at distinct points, and is positive or negative on the intervals in between. A negative value says that the system is unstable to a change in the boundary conditions, suggesting that the true ground state of the model in the thermodynamic limit is incommensurate with the structure of the finite cluster being used. A positive value reveals a stable ground state, and can sometimes be used with finite-size scaling to extract the behavior of the stiffness in the thermodynamic limit. This is in particular the case in the Néel and collinear regions.

The spin stiffness for the unfrustrated spin-$\frac{1}{2}$ Heisenberg model on a 2D square lattice has been calculated directly by series expansion [2], $\rho_s = (0.18 \pm 0.01)J_1$, and by second-order spin-wave theory (SSWT) [3], $\rho_s = (0.181 \pm 0.001)J_1$. Furthermore, the spin-wave velocity $c$ and the transverse susceptibility $\chi_\perp$ have also been calculated in SSWT, and since the ensemble of values fulfill the hydrodynamic relation [1] $\rho_s = c^2 \chi_\perp$ to a good approximation, there is strong evidence for the accuracy of the SSWT values [3]. However, a previous attempt to extract the value of $\rho_s$ from exact diagonalizations (ED) yielded $\rho_s = 0.125 J_1$ [4], which is far away from the other results. This is not too bothering regarding that the ED value of $\rho_s$ (and $c$) was obtained from the correction terms in the finite-size scaling analysis and as such looses accuracy due to cancelations, and is further influenced by higher-order corrections which are not known. To obtain more accurate values of the spin stiffness, we here set out to calculate the spin stiffness directly by using EDs to evaluate $\rho_s$ as a correlation function. In contrast to two recent works which have employed ED and finite twists on the square and triangular lattices [5,6], our method preserves more symmetries, and we can treat clusters of up to 36 sites.

By performing a careful finite-size extrapolation we arrive at a value of the stiffness in the non-frustrated case, $\rho_s = (0.183 \pm 0.003)J_1$, in excellent agreement with the SSWT and series-expansion results. In the case of frustrating interactions, things are more complicated. In a previous ED study [4], the order-parameter was found to vanish in the region $0.34 \lesssim J_2/J_1 \lesssim 0.68$, and one of our aims was to find out whether a direct calculation of the stiffness would corroborate this result. Our results suggest that the stiffness vanishes for $0.4 \lesssim J_2/J_1 \lesssim 0.6$, but there is also a tendency of the stiffness to blow up in the region $J_2/J_1 \lesssim \frac{1}{2}$. A similar tendency is found in a first-order SWT (FSWT). In the latter case, this burst is a signature of the breakdown of SWT as $J_2/J_1$ approaches the classical transition point $J_2/J_1 = \frac{1}{2}$.

We start with the general Heisenberg Hamiltonian

$$H_0 = \sum_{(i,j)} J_{ij} \left[ \frac{1}{2} \left( S_i^+ S_j^- + S_i^- S_j^+ \right) + S_i^z S_j^z \right] , \qquad (2)$$

where the sum goes over all pairs of sites $(i,j)$, and introduce a local rotation at site $i$ by $\theta_i$ around the $z$-axis, $S_i^+ \to S_i^+ e^{+i\theta_i}$, $S_i^- \to S_i^- e^{-i\theta_i}$, and $S_i^z \to S_i^z$, so that $S^z_{\text{tot}}$ is unchanged. A Mac-Laurin expansion around $\theta_{ij} \equiv \theta_i - \theta_j = 0$ gives to order $\theta_{ij}^2$ [5]

$$H = H_0 + \sum_{(i,j)} \left[ \theta_{ij} j^{(s)}_{ij} - \frac{1}{2}\theta_{ij}^2 T_{ij} \right],  \quad (3)$$

where $j^{(s)}_{ij} = \frac{i}{2} J_{ij}(S^+_i S^-_j - S^-_i S^+_j)$ is the z-component of the spin-current operator, and $T_{ij} = \frac{1}{2}J_{ij}(S^+_i S^-_j + S^-_i S^+_j)$ is the "spin-kinetic-energy" operator. To obtain the spin stiffness, a uniform twist $\theta$ is introduced between each pair of adjacent rows, i.e., $\theta_{ij} = \theta[(r_i - r_j) \cdot \hat{y}]$, and to second order in $\theta$ one has $\langle H(\theta) \rangle = \langle H_0 \rangle + \frac{1}{2}N\theta^2 \rho_s$. This gives a direct expression for $\rho_s$, which for the $J_1$-$J_2$ model [$J_{ij} = J_1$ ($J_2$) for nearest (next-nearest) neighbors] reads

$$\rho_s = \frac{2}{N} \left[ \frac{1}{2}\langle -T_y(s) \rangle + \langle 0| j^{(s)}_y P_0 \frac{1}{E_0 - H} P_0 j^{(s)}_y |0\rangle \right] \equiv \text{TY} + \text{JY}, \quad (4)$$

$$T_y(s) = \frac{1}{2} \sum_i \left[ J_1 S^+_i S^-_{i+\hat{y}} + J_2(S^+_i S^-_{i+\hat{x}+\hat{y}} + S^+_i S^-_{i-\hat{x}+\hat{y}}) + \text{h.c.} \right], \quad (5)$$

$$j^{(s)}_y = \frac{i}{2} \sum_i \left[ J_1 S^+_i S^-_{i+\hat{y}} + J_2(S^+_i S^-_{i+\hat{x}+\hat{y}} + S^+_i S^-_{i-\hat{x}+\hat{y}}) - \text{h.c.} \right], \quad (6)$$

where the JY terms comes from second order perturbation theory, and where $P_0 = 1 - |0\rangle\langle 0|$ is the projection operator projecting on the space orthogonal to the ground state. Note that the $J_2$ term has two terms per site and that the expectation values are evaluated in the *non*-twisted space. The stiffness is now expressed as a sum of a "kinetic-energy" term TY, which is easy to calculate, and a spin-current spin-current correlation function JY, which needs some computational efforts to be evaluated.

To calculate JY, we use a continued-fraction expansion [7] where we repeatedly apply the Hamiltonian on the spin-current state $|f_0\rangle \equiv P_0 j^{(s)}_y |0\rangle$, which is antisymmetric under spin inversion and under reflection on the $x$-axis. The loose of diagonal reflection symmetry implies a doubling of the size of the Hilbert space, which for the 36-site cluster is now $\sim 3 \times 10^7$. The expansion normally converges very quickly and $\rho_s$ is obtained with five significant digits after five to ten iterations.

As a test of our method, we first considered the ferromagnetic model, where both $J_1$ and $J_2$ are negative. The ferromagnetic state with $S^z_{\text{tot}} = 0$ is the symmetric superposition of all $S^z_{\text{tot}} = 0$ states. The transverse correlations are easily obtained as $\langle \frac{1}{2}(S^+_i S^-_j + S^-_i S^+_j) \rangle = \frac{1}{4}N/(N-1)$ and, for periodic boundary conditions, the JY term is identically zero. The order parameter lies in the $z = 0$ plane, and one is really measuring the (transverse) spin stiffness (compare the AF case below), $\rho_s = \frac{1}{4}(J_1 + 2J_2)N/(N-1)$. This result is exactly reproduced in our exact diagonalizations.

The antiferromagnetic case differs from the FM case both by the necessity to consider the spin-current term and by the ground state being rotationally invariant. The latter fact means that the twists are not orthogonal to the order parameter, but instead we calculate the rotational average of the stiffness. Since the stiffness for a twist around the Néel (or the collinear) order parameter is zero, we have to multiply our result by a factor $\frac{3}{2}$ to arrive at the ordinary transverse stiffness. Let us first consider the unfrustrated case.

To extract the values of thermodynamic quantities from finite-size calculations it is of crucial importance to have good knowledge about the scaling behavior of the quantities of interest. A great deal of information can be obtained from studying how the spin-wave theory behaves under scaling, or from the finite-size analysis of the non-linear $\sigma$ model [8]. The FSWT expression for the stiffness [3] can be written as

$$\rho_s = -\frac{E_0}{2N} + \frac{JS}{2}\frac{2}{N}\sum_k \left( \epsilon_k - \frac{1}{\epsilon_k} \right), \quad (7)$$

where $E_0$ is the LSWT ground-state energy and $\epsilon_k$ is related to the LSWT dispersion relation by $\omega_k = 4SJ\epsilon_k$. By looking at the $k$-sums involved, one finds that the correction to the ground state energy per site $E_0/N$ scales as $N^{-3/2}$ and that the correction to the second term scales as $N^{-1/2}$. Using the rotational invariance of the ground state, we can further rewrite the ED expression (4) as

$$\rho_s = \frac{3}{2}\left[ -\frac{E_0}{3N} + \text{JY} \right]. \quad (8)$$

The physical content of the first term is thus exactly the same in the both cases, and it is known that the correction to $E_0/N$ goes as $N^{-3/2}$ also in the ED case [8]. It is therefore wise to use the same scaling as in SWT also for the JY term, $\text{JY}_N - \text{JY}_\infty \propto N^{-1/2}$. With these scaling laws we can extrapolate the TY and JY terms separately, and then finally obtain the stiffness in the thermodynamic limit as

$$\rho_{s,\infty} = \text{TY}_\infty + \text{JY}_\infty. \quad (9)$$

As was noted in Ref. [4], the extrapolated value is sensitive to which set of cluster sizes one uses. In Tab. I, the results for the different clusters are presented and in Tab. II, the results of the various extrapolations are presented together with error bounds coming from a $\chi^2$-fit of the values to a straight line. As seen in Tab. II, the set of clusters with $\{18, 20, 32, 36\}$ sites gives the best result in the non-frustrated case. When turning on $J_2$ we are in a much less understood regime. Semi-classically, there is a sharp transition at $J_2/J_1 = \frac{1}{2}$, from a Néel state to a collinear state. However, going to $S = \frac{1}{2}$, there may well be a widening of the transition region and a region with a spin-liquid ground state may open up. Indeed, the earlier finite-size studies suggested that the Néel and collinear states are separated by an intermediate region for $0.34 \lesssim J_2/J_1 \lesssim 0.68$ [4]. On the other hand, besides a number of works which have also found a reduced Néel stability, the large-$S$ studies using Schwinger-boson mean-field theory [9] or a self-consistent spin-wave theory [10] show an increased Néel stability with respect to the classical case. Since these methods are only trust-worthy for large values of $S$, the discrepancy for $S = \frac{1}{2}$ is not necessarily significant. It is also not surprising that a self-consistent mean-field calculation of $\rho_s$ yielded a stiffness which does not vanish until $J_2/J_1 \sim 0.6$ [11].

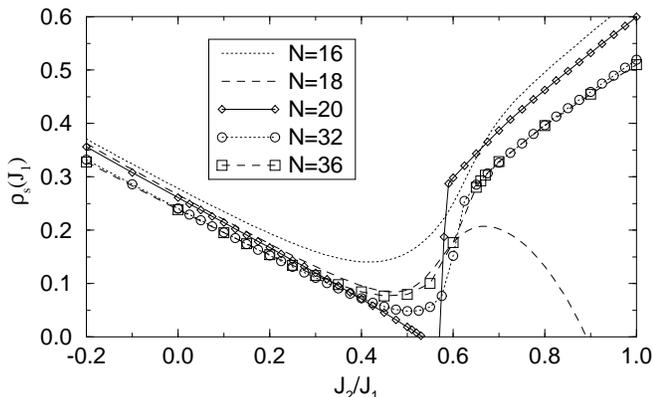

FIG. 1. The stiffness $\rho_s$ for the various clusters being used. The 18-site cluster shows a negative stiffness for big $J_2$, and the 20-site cluster has a change in the ground-state symmetry around $J_2/J_1 = 0.58$.

A good test of our numerical program is to consider the limit $J_2/J_1 = \infty$, or $J_1 = 0$, $J_2 = 1$. Here, the two sublattices decouple and the energy (stiffness) should be twice the energy (stiffness) of the subclusters. This is indeed exactly what we obtain. As $J_2/J_1$ increases, we thus expect to see a decrease in the stiffness followed by an increase as the two sublattices become individually ordered. The minimum should be somewhere not too far from the classical break point $J_2/J_1 = \frac{1}{2}$. For the 18-site cluster the stiffness should go negative for large $J_2/J_1$ because that cluster is not compatible with the structure of two antiferromagnetic sublattices. These observations agree with the results for the finite clusters presented in Fig. 1.

Unfortunately, the individual properties of the clusters result in rather strong peculiarities. In the 20 and 36-site cases, there is a change in the symmetry of the ground state as the two sublattices become individually antiferromagnetically ordered. In the 20-site case, this causes an abrupt jump in the stiffness, while in the 36-site case the transition is very smooth.

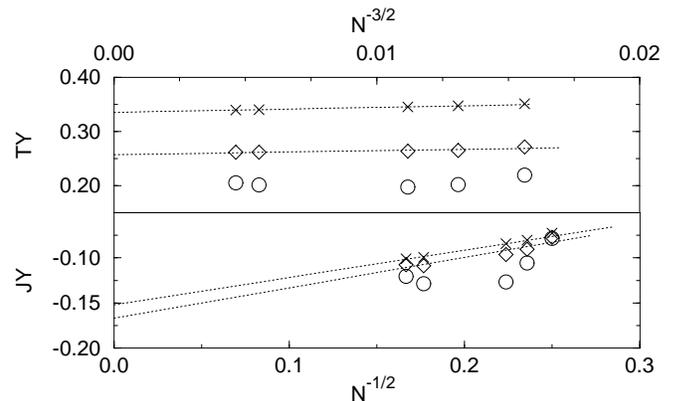

FIG. 2. The finite-size data for TY and JY for $J_2/J_1 = 0$ (crosses), 0.2 (diamonds), and 0.4 (circles) together with fits for $J_2/J_1 = 0.0$ and 0.2. For small frustration, $J_2/J_1 < 0.3$, the scaling law is well satisfied, but in the intermediate region the results do not line up.

Given the strong individual differences in Fig. 1, it is not evident how to extrapolate to $N = \infty$ for the various degrees of frustration. In Fig. 2, we show the actual data which we try to fit with our scaling laws, for $J_2/J_1 = 0$, 0.2, and 0.4. In the region $0.3 \lesssim J_2/J_1 \lesssim 0.6$, the results for JY do not line up and the extrapolation to $N = \infty$ is unreliable. In Fig. 3, we show the results of extrapolations using a few different sets of clusters. In the intermediate region our results are scattered. The FSWT result is obtained by generalizing Eq. (7).

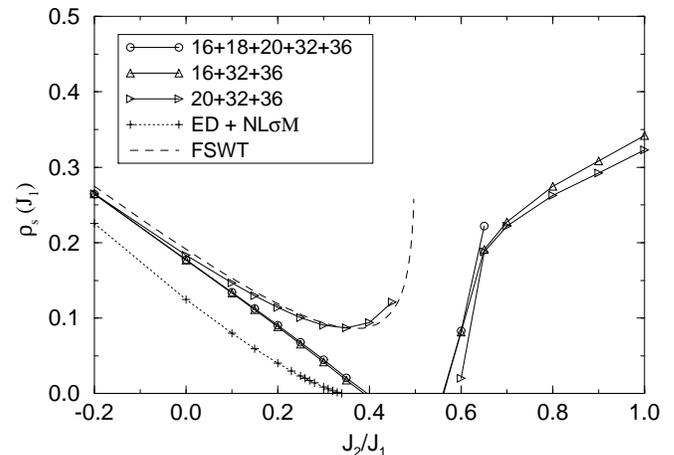

FIG. 3. The extrapolated value of the stiffness for some choices of clusters together with earlier ED results (ED+NL$\sigma$M) and FSWT.

By excluding the 20-site cluster, the results suggest a vanishing LRMO in the region $0.4 \lesssim J_2/J_1 \lesssim 0.6$ in rather good agreement with the previous ED results [4] (where the stiffness vanished at the same point as the order parameter). The extrapolation from the $\{20, 32, 36\}$- site clusters follows the FSWT result closely, but the derivative of the latter diverges as $J_2/J_1 \to \frac{1}{2}$ and we consider this similarity to be fortuitous. If one really were in the Néel regime all the way to $J_1/J_2 \lesssim \frac{1}{2}$, the coupling constant in the non-linear $\sigma$ model, $g \propto c/\rho_s$, would be roughly constant over the entire region and there is no reason why the finite-size scaling should cease to be valid. This is however the case as seen in Fig. 2, and we conclude that the intermediate region has neither Néel nor collinear order, and that a first-order transition from Néel to collinear order as suggested in Refs. [9] and [10] is inconsistent with this result.

Since we consider our result from the $\{16, 32, 36\}$-cluster extrapolation to be good, we can combine it with the previous ED results for the transverse susceptibility $\chi_\perp$ [4] to obtain the spin-wave velocity $c$ from the hydrodynamic relation $c = \sqrt{\rho_s/\chi_\perp}$. The result is shown in Fig. 4. The result is in fair agreement with LSWT, $c = J_1\sqrt{2(1-2J_2/J_1)}$, but close to the phase boundary the result may not be trusted since the susceptibility and the stiffness do not vanish at the same point. In the non-frustrated case, our best value, $\rho_s = 0.183J_1$, yields $c = 1.67J_1$ in excellent agreement with the SSWT result [12,3] $c = 1.664J_1$.

Bonča et al. [5] have reported results for $\rho_s$ for the 16 and 20-site clusters. Their results differ from ours due to a number of lapses on their side. First of all, they did not include the $J_2$ terms in Eqs. (5,6). Secondly, they missed the factor $\frac{3}{2}$, which compensates for the rotational symmetry of the ground state, and finally they did not use the proper power laws in their extrapolation to the thermodynamic limit.

It would be of great interest to extract some precise signature of the ground state in the intermediate region. This is however not possible from the spin stiffness. Even a spin liquid may have a finite stiffness for a finite system and in the region where the finite-size scaling does not work, we can only exclude Néel and collinear LRO. Our results strongly suggest the existence of an unconventional ground state in a wide intermediate region, but its nature has to be revealed by a more detailed examination of the correlation functions.

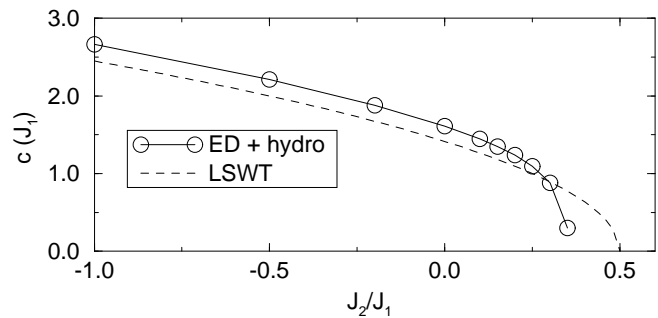

FIG. 4. The spin-wave velocity obtained by using the hydrodynamic relation. As a comparison, the LSWT results are shown.

We thank P. Lecheminant for illuminating discussions and IDRIS (France) for the computing time made available on their Cray C98. The work of T.E. was financed by the Swedish Natural Science Research Council.

TABLE I. Values of $\rho_s$, TY, and JY for finite clusters and no frustration.

| N | TY | JY | $\rho_s$ |
|---|---|---|---|
| 16 | 0.35089 | −0.07248 | 0.27841 |
| 18 | 0.34699 | −0.08054 | 0.26646 |
| 20 | 0.34540 | −0.08433 | 0.26107 |
| 32 | 0.34009 | −0.09938 | 0.24071 |
| 36 | 0.33943 | −0.10084 | 0.23859 |

TABLE II. Extrapolated values for TY, JY, and $\rho_s$ for $J_2 = 0$, with the uncertainty in the last digit given in parentheses.

| Cluster sets | $TY_\infty$ | $JY_\infty$ | $\rho_{s,\infty}$ |
|---|---|---|---|
| 16,18,20,32,36 | 0.3345(7) | −0.157(5) | 0.177(6) |
| 16,20,32,36 | 0.3344(6) | −0.159(6) | 0.176(7) |
| 16,32,36 | 0.3344(2) | −0.160(5) | 0.174(5) |
| 18,20,32,36 | 0.3352(1) | −0.152(3) | 0.183(3) |
| 20,32,36 | 0.3351(2) | −0.152(5) | 0.184(5) |